\documentclass[aps,pra,showpacs,
twocolumn,groupedaddress]{revtex4}
\usepackage{graphicx}
\usepackage{epsfig}
\usepackage{amsmath}
\usepackage{amssymb}
\newcommand{\tr}{\mathrm{tr}}
\newcommand{\OO}{\mathcal{O}}
\newcommand{\hA}{{A}}

\newcommand{\Eqref}[1]{Eq.~(\ref{#1})}

\def\ket#1{\left| #1\right>}

\newcommand{\mr}[1]{\mathrm{#1}}

\newcommand{\up}{\hspace{-3 pt}\uparrow}
\newcommand{\dn}{\hspace{-3pt}\downarrow}

\newcommand{\bS}{\mathbf{S}}
\newcommand{\bA}{\mathbf{A}}

\begin{document}

\title{Quantum Description of Nuclear Spin Cooling in a Quantum Dot}
\author{H. Christ}
\author{J. I. Cirac}
\author{G. Giedke}
\affiliation{ Max--Planck--Institut f\"{u}r Quantenoptik,
Hans-Kopfermann--Str. 1, D--85748 Garching, Germany }
\date{\today}

\begin{abstract}
  We study theoretically the cooling of an ensemble of nuclear spins coupled
  to the spin of a localized electron in a quantum dot. We obtain a master
  equation for the state of the nuclear spins interacting with a sequence of
  polarized electrons that allows us to study quantitatively the cooling
  process including the effect of nuclear spin coherences, which can lead to
  ``dark states'' of the nuclear system in which further cooling is inhibited.
  We show that the inhomogeneous Knight field mitigates this effect strongly
  and that the remaining dark state limitations can be overcome by very few
  shifts of the electron wave function, allowing for cooling far beyond the dark state limit.
  Numerical integration of the master equation indicates, that
  polarizations larger than 90\% can be achieved within a millisecond
  timescale.
\end{abstract}

\maketitle

\section{Introduction}\label{sec:intro}
Nuclear spins are one of the best studied quantum systems and highly developed
techniques such as NMR have allowed detailed study of properties and dynamics
of molecular and solid state systems \cite{Abr61}. Due to their very
long decoherence time nuclear spins (and hyperfine levels) have also played a
central role in many approaches to the implementation of quantum information
processing (QIP) \cite{CiZo95,GeCh97,CFH97,Kan98,WKN00}.

Recently, the localized ensemble of nuclear spins in a quantum dot
(QD) has received special attention in the context of QIP with
electron spins in QDs: the nuclei couple via a Fermi contact
interaction to the electron spin \cite{SKL03} and, as predicted by
theory \cite{BLD99,ENF01,KLG02,MER02,CoLo04}, have been shown in
recent experiments to constitute the major source of decoherence
of electron spin qubits in some of the most promising QD-based
implementations \cite{PJT+05,KFE+05}. The vice of this strong
coupling is turned into a virtue when the electron is used to
manipulate the state of the nuclear ensemble. This has long been
exploited in dynamical nuclear polarization (DNP)
\cite{Ov53,Lam68,MeZh84,SAOO01} in bulk systems and afforded many
insights in the spin dynamics in solids \cite{PLSS77,MeZh84}.

DNP \emph{in quantum dots} has come into focus more recently in
the context of QIP, since strongly polarized nuclei could lead to
much longer electron spin dephasing times \cite{CoLo04}, provide
strong local magnetic field gradients required in quantum
information proposals \cite{TED+05,CL06}, and even allow to
utilize the nuclear spins themselves as long-lived quantum memory
\cite{TIL03,TGC+04}. More generally, a highly polarized nuclear
spin ensemble in a QD provides, together with the electron spin, a
strongly coupled, well isolated mesoscopic quantum systems with
close similarities to the Jaynes-Cummings model in quantum optics
\cite{TGC+04,SZSS05,CGT+unpub}, with the fully polarized state
corresponding to the vacuum in all cavity modes. Thus ultra-high
DNP in QDs may open the door to realize cavity-QED in quantum dots
and implement tasks such as state engineering.

Experimentally, significant nuclear polarization in self-assembled
QDs has been achieved \cite{BSG+05,EKL+06,GOV+05,LMBI06,AFH06}.
However, the degree of polarization in these experiments was still
too low to improve electron spin coherence times considerably and
still far from the ground state.

Theoretically, cooling dynamics has mostly been considered in the
spin temperature approximation \cite{Abr61,MeZh84,DeHu05b,RuLe06},
in which coherences among the nuclear spins are neglected.  This
is appropriate if, as in bulk or quantum well systems, there is no
fixed electron wave function and many motional states are
involved, or if the nuclear dephasing rate is large.  In quantum
dots, however, the nuclei interact collectively with an electron
in the motional ground state of the QD and the higher motional
levels are far detuned.  Therefore the coupling strength of each
nucleus is fixed, and well defined phase relationships between the
nuclear spins can build up, necessitating a quantum treatment of
the process, which was first pointed out by Imamo\u{g}lu \emph{et
al.} \cite{IKTZ03}, who showed that the cooling process can be
inhibited by so-called dark states, which trap excitations and
potentially result in serious constraints on the achievable
polarizations. While it was pointed out in \cite{IKTZ03} that
inhomogeneities (either inherent in the system or introduced
actively by modulating the wave function of the electron) can
mitigate this problem, these ideas were put to numerical test only
in very small 1D systems of 10 nuclear spins. However, the effect
of inhomogeneities is expected to be reduced for realistic larger
systems \cite{TIL03}, and thus limitations due to dark states are
more severe \footnote{In a fully homogeneous system only a
fraction of $\OO(1/\sqrt{N})$ spins can be cooled before the
system is trapped in dark states.}.

We consider the cooling of $N$ nuclear spins in a QD through
interaction with polarized electrons. One cooling cycle consists
of ($a$) initialization of the electron spin in a well-defined
direction, and ($b$) evolution of the combined system for a
``short'' time. In this way the electron spin acts effectively as
a $T=0\,$-reservoir for the nuclear spin bath, and pumps
excitation out of it.

We derive in a consistent manner a full quantum model of this
process, which allows us to numerically study particle numbers of
up to $N \sim 10^3$. We show that a sufficient inhomogeneity of
the couplings leads to a dephasing of nuclear spin states and thus
limitations due to dark states are partially lifted. We
demonstrate that enhanced cooling protocols involving only a few
($\leq 10$) modulations of the electron wave function, allow to
fully overcome these limitations, indicating that Overhauser
fields above $90\%$ of the maximal value can be created within the
nuclear spin diffusion time.

The paper is organized as follows: In Sec.~\ref{sec:model} we present
the generic cooling protocol and analyze its performance in
Sec.~\ref{sec:numerical}; the applicability of the scheme to some
specific physical systems is studied in Sec.~\ref{sec:adapt}.

\section{The cooling scheme}\label{sec:model}
\emph{Interaction-- }
The Fermi contact interaction between an ($s$-type conduction band) electron
spin ${\mathbf{S}}$ and the spins $\mathbf{I}_i$ of the lattice nuclei leads
to a Heisenberg like coupling \mbox{$A \alpha_i {\mathbf{I}}_i \cdot
  {\mathbf{S}}$} to the nuclear spin at lattice site $i$, where $A$ sets the
overall strength of the hyperfine interaction and the factor
$0<\alpha_i<1$ is determined by the probability to find the electron
at site $i$ and the gyromagnetic ratio of the $i$th nucleus
\cite{SKL03}. In the presence of an external magnetic field $B_{\rm
ext}$ we write the Hamiltonian of the spin system with the
collective nuclear spin operators $A^\mu = \sum_i g_i I^\mu_i$
($\mu=\pm,z$) as ($\hbar =1$)
\begin{equation}
{H}=\frac{g}{2}\left(\hA^+{S}^- + {S}^+\hA^-\right) + g
\hA^z{S}^z+g^*\mu_BB_{\rm ext}  S^z,\label{Hamiltonian}
\end{equation}
where we have defined $g=A\sqrt{\sum_i \alpha_i^2}$ and $g_i =
\alpha_i/\sqrt{\sum_i \alpha_i^2}$, such that $\sum_i g_i^2 =1$, and
denoted  the electron $g$-factor by $g^\ast$ and the Bohr magneton
by $\mu_B$.

We do not consider the Zeeman energy of the nuclear spins, because
for typical QDs it is much ($10^3$ times) smaller than the
electron's Zeeman energy~\cite{SKL03}, and similarly we neglect
the even smaller dipolar interaction between the nuclei.  The
effects of these are briefly discussed at the end of
Sec.~\ref{sec:numerical}. Finally, we restrict the analysis to
nuclear spins $I=1/2$ and one nuclear species only in this
article.

The first part of the above Hamiltonian exchanges spin excitation between the
electron and the nuclei, and it is this mechanism that is used to create
polarization. The second part of the Hamiltonian constitutes a ``quantum''
magnetic field, the Overhauser field, for the electron spin generated by the
nuclei.

\emph{The cooling scheme-- } We assume initially the electron spin
to be pointing in the $-z$-direction $|\psi_{e^-}\rangle =
|\dn\rangle$. In the absence of a magnetic field this initial
state defines the axis of quantization. The cooling cycle we
consider is an iteration between evolution with
Hamiltonian~Eq.(\ref{Hamiltonian}), and reinitialization of the
electron to $|\dn\rangle$. The nuclei effectively ``see'' a large
cold reservoir of electron spins and the concatenated evolution of
the nuclear spin density matrix becomes
\begin{equation}
\label{BasicEvolution} \rho \rightarrow \ldots U_t \tr_e \left[U_t
\Big( \rho \otimes | \dn \rangle \langle \downarrow
\hspace{-0.1cm} | \Big) U_t^\dag \right] \otimes  | \dn \rangle
\langle \downarrow \hspace{-0.1cm}  | U_t^\dag\ldots\,\,\, .
\end{equation}
Here $U_t = \exp(-i H t)$ is the time evolution operator, ${\rm
tr}_e$ denotes the trace over the electron, and here and in the
following $\rho$ will denote the state of the nuclear spin system
only. Spin polarized currents or optical pumping with polarized
light give rise to a polarized electron bath, but also the fast
electrical control available in double QDs~\cite{PJT+05} allows
for the creation of nuclear spin polarization without the need for
pre-prepared electrons, as we will detail in the last section of
this article.

Considering small times for the evolution in each individual step
of the cooling protocol, we expand the time evolution operators in
Eq.(\ref{BasicEvolution}) to second order. The standard deviation
of the $A^{\pm,z}$-terms scales as $ A \sqrt{\sum_i \alpha_i^2} =
g \sim \OO(A/\sqrt{N})$ for the initially totally mixed nuclear
spin state, and thus for $\Delta t \ll g^{-1} \sim \sqrt{N}/A$ we
neglect higher orders. The readily obtained master equation
\begin{align}
\rho&_ {t+\Delta t}  -\rho_t=  i \frac{g \Delta t}{2}
[\hA^z,\rho_t]
-\frac{g^2 (\Delta t)^2}{8}[\hA^z,[\hA^z,\rho_t]\nonumber\\
& - \frac{g^2 (\Delta t)^2}{8} \left( \hA^+ \hA^- \rho_t + \rho_t
\hA^+ \hA^- -2 \hA^- \rho_t \hA^+ \right),\label{Meq}
\end{align}
contains a Hamiltonian part arising from the Overhauser field and a
contribution in Lindblad form. The latter generates the nuclear spin
polarization, and has been studied in the limit of homogeneous coupling
constants in the context of superradiance~\cite{BSH71,GrHa82,AEI93}.

As polarization builds up and $g\langle A^z\rangle\gg A/\sqrt{N}$ the
Hamiltonian terms on the right hand side of Eq.(\ref{Meq}) may become large
(for fixed time step $\Delta t$). To preserve validity of the master equation
one can either reduce the interaction time $\Delta t < A^{-1}$ or assume that
the Overhauser field $\langle A^z\rangle$ is approximately compensated by an
applied magnetic field, so that $\langle gA^z -g^*\mu_B
B_\mr{ext}\rangle\Delta t\ll1$ for all times. In the latter case $\Delta t$ is
short enough to ensure quasi-resonant hyperfine flips despite the random
detunings stemming from the fluctuating Overhauser field and at the same time
large enough to guarantee a fast cooling rate \footnote{Ensuring the validity
  of \Eqref{Meq} for all times by retuning $B_\mr{ext}$ assumes that the
  standard deviation of $gA^z$ remains bounded by $\OO(A/\sqrt{N})$ thus
  keeping the error in each cooling step small. Computing $\mr{Var}A^z$ in
  each step and choosing $\Delta t$ accordingly guarantees correctness. In
  general, the polarizing process is expected to \emph{decrease} $\mr{Var}A^z$
  from the initial value in the maximally mixed state. This is confirmed by
  exact numerical calculations for small particle numbers. We are confident
  that this holds for large $N$, too, since the generic states exhibit
  standard deviation $\le \OO(A/\sqrt{N})$ (as evidenced by the variance of
  the maximally mixed state). Moreover, the standard deviation in the maximal
  entropy state of total polarization $P$ is $\OO((1-P^2)A/\sqrt{N})$ for all
  $P$. Similar reasoning holds for the $x$- and $y$-directions.}.  This is the
situation we investigate in the following. Without retuning the
system in this manner the polarization rate becomes dependent on
the polarization itself and the emerging non-linearities give rise
to the bistability effects observed in
\cite{OnTa04,HWH+04,KFE+05,AFH06,BUA+06,MLBI07,TWR+07}and limit
the final polarization.

\emph{Homogeneous Coupling-- } Before we discuss general
inhomogeneous couplings, consider for a moment the homogeneous case,
$\alpha_i\propto 1/N$, as a demonstration of some interesting
features of the above master equation. In this case, the operators
$A_{\pm,z}$ appearing in \Eqref{Meq} form a spin algebra $I_{\pm,z}$
and the collective angular momentum states (Dicke states)
$\ket{I,m_I,\beta}$ provide an efficient description of the system
dynamics \cite{ACGT72,TIL03}: the total spin quantum number $I$ is
not changed by $A_{\pm,z}$ and the effect of \Eqref{Meq} is simply
to lower (at an $(I,m_I)$-dependent rate) the $I_z$ quantum number.
If $m_I=-I$ is reached, the system can not be cooled any further,
even if (for $I\ll N/2$) it is far from being fully polarized. These
dark states~\cite{IKTZ03,TIL03} are a consequence of the collective
interaction \Eqref{Hamiltonian}. Thus spin excitations are trapped
and cooling to the ground state prevented. We evaluate the steady
state polarization $\langle {I}^z \rangle_{\rm ss}=\langle \sum_i
{I}_i^z /\sqrt{N} \rangle_{\rm ss}$ as
\begin{equation}\label{HomCool}
\frac{\langle I^z \rangle_{\rm ss}}{\langle I^z
\rangle_0}=\frac{2}{2^N N}\sum_{I=0}^{N/2} I(2I+1) D_I =
\sqrt{\frac{8}{\pi N}} + \OO(1/N),
\end{equation}
i.e. for a mesoscopic number of particles the obtained
polarization is negligible. In the above equation $\langle I^z
\rangle_0$ is the expectation value in the completely polarized
state, $D_I= \left(
\begin{array}{c}
  N\\
  N/2-I\\
\end{array}\right)  - \left(
\begin{array}{c}
  N\\
  N/2-I-1\\
\end{array}\right)$ is the degeneracy of the subspaces of
different total angular momentum, and the last equality has been obtained by
employing the Stirling formula.

\begin{figure}
\begin{center}
\includegraphics[width=\columnwidth,height=3.9cm]{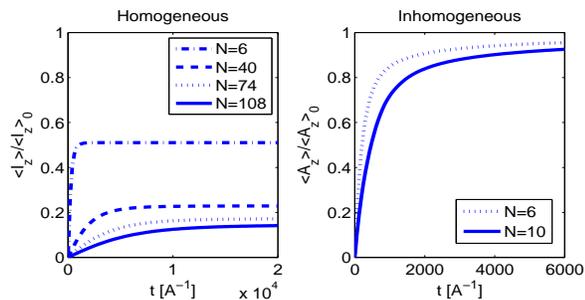}
 \caption{(Color online) Exact polarization dynamics.
Left: Homogeneous case, $g_j=1/\sqrt{N}$. Right: In the
inhomogeneous case, $g_j \propto \exp(-(j-N/2-1/4)^2/w^2)$. The
term 1/4 is added to account for asymmetry between electron wave
function and the lattice and avoid symmetry effects for this small
scale system.} \label{fig1}
\end{center}
\end{figure}

Evolving the nuclei according to Eq.(\ref{Meq}), we find the exact
time evolution of the polarization as shown in Fig.~\ref{fig1}. In
these and the following simulations $g \Delta t =0.1$, i.e. $\Delta
t =0.1 g^{-1} \sim 0.1 \sqrt{N}/A$. As expected the polarization
decreases as $1/\sqrt{N}$ as $N$ increases, which underlines the
importance of the nuclear spin coherences. In particular this shows
that an incoherent spin temperature description of the process would
give even qualitatively wrong results. The timescale over which the
steady state is reached is $\sim N/ (g \Delta t A)$.

\emph{Inhomogeneous Coupling-- } Consider now an inhomogeneous
wave function. The results for the exact evolution of the quantity
of interest, $\langle A^z \rangle$, are shown in Fig.~\ref{fig1}.
The coupling constants $g_j$ in this example are taken from a 1D
Gaussian distribution with width $N/4$.\footnote{In this article
we focus on Gaussian electron wave functions, which approximate
experimental conditions well. For the coherent phenomena we
discuss, the distribution of the groups of similarly coupled spins
is of major importance. This property is generally mainly
determined by the width and dimensionality of the wave function,
and only to a smaller extent by its exact functional form. } The
most important and striking feature is that in this situation
almost complete polarization is obtained.

The reason that this is possible here is \textit{not} that there
are no dark states in the case of inhomogeneous coupling
constants. On the contrary it has been shown that there exists an
one-to-one mapping~\cite{TIL03} from the familiar homogeneous dark
states ($|I,-I,\beta\rangle$ in the Dicke basis) to their
inhomogeneous counterparts, defined by $A^-|D\rangle=0$. The
reason for obtaining high polarization beyond the homogeneous
limit is the Hamiltonian part of the master equation~(\ref{Meq}).
To illustrate this point, consider two spins with coupling
constants $g_1 \neq g_2$. Then the dark state $|\Psi_D\rangle
\propto g_2|\up \downarrow\rangle - g_1 |\dn \uparrow\rangle$
evolves due to the $A^z$-term in Eq.(\ref{Meq}) to $e^{i \delta_g
t} g_2|\up \downarrow\rangle - e^{- i \delta_g t} g_1 |\dn
\uparrow\rangle$, where $\delta_g$ is proportional to $g_1 -g_2$.
Obviously this state will become ``bright'' again after a time
$\propto1/|g_i-g_j|$ and ${A}^-|D\rangle \neq 0$.  This process is
first order and, as we will detail later, ``delivers'' coolable
excitations sufficiently fast to maintain a high cooling rate.

\section{Polarization Dynamics}\label{sec:numerical}
The polarization dynamics of the nuclear ensemble is governed by \Eqref{Meq}.
While for homogeneous systems the collective angular momentum Dicke basis
enables an efficient description of the problem, for realistic large and
inhomogeneous systems more effort is required.

To study the evolution of the nuclear polarization, we are interested in the
individual spin expectation values $\langle\sigma_i^+\sigma_i^-\rangle$. These
depend, via \Eqref{Meq} on all the elements of the covariance matrix
\begin{equation*}
\gamma_{ij}=\langle \sigma_i^+ \sigma_j^- \rangle,
\end{equation*}
which, in turn, depend on higher order correlations as seen from the equations
of motion
\begin{align}
\label{eqmgamma} \frac{\Delta \gamma_{ij}}{\Delta t} =  \xi_{ij}
\gamma_{ij} - \kappa  \sum_k g_k \Big( & - g_i \langle \sigma_k^+
[\sigma_i^+, \sigma_i^-] \sigma_j^-  \rangle \nonumber \\   + &
g_j \langle \sigma_i^+ [\sigma_j^-, \sigma_j^+] \sigma_k^-
\rangle\Big),
\end{align}
where $\xi_{ij}=i g (g_j -g_i)/2  - g^2 \Delta t (g_j -g_i)^2 /8  $
and $\kappa = g^2 \Delta t/8$ and the $\sigma^\mu_i$ refer to the Pauli
matrices at site $i$.

The simultaneous solution of the ensuing hierarchy of equations is only
feasible for very small particle numbers $N$ and further approximations are
needed to treat the large systems of interest. We introduce several ways,
labeled ($i$) to ($v$), of closing this set of equations and discuss their
validity and implications in detail below.

In the strongest approximation ($i$) all coherences between
different spins are neglected yielding independent rate equations
for each individual nuclear spin. This reproduces essentially the
spin-temperature description commonly employed in the discussion
of bulk DNP \cite{Abr61,MeZh84} (each subset of spins with
identical coupling strengths $g_i$ is assigned its own effective
temperature).  This approach cannot reproduce the quantum effects
we want to study, but it can serve as a benchmark for how strongly
these are influencing the cooling process.

The simplest approximations that take quantum coherences between nuclear spins
into account close the hierarchy of equations at the level of second order
correlations. Our approximation ($ii$) is motivated by the generalized
Holstein-Primakoff description \cite{HoPr40}, which in lowest order treats the
nuclei as bosonic modes $\sigma_i^-\to a_i$. The bosonic commutations
relations $[a_i,a_j^\dagger]=\delta_{ij}$ yield a closed set of equations for
the elements of the covariance matrix $\gamma$. The bosonic description is
known to be accurate for highly polarized and moderately inhomogeneous systems
\cite{CGT+unpub} and allows to bring results and intuition from quantum optics
to bear in the spin system discussed here.  Dark states are included in the
form of the vacuum of the collective mode $b=\sum_ig_ia_i$ coupled to the
electron in \Eqref{Hamiltonian}. For unpolarized systems (with on average
$1/2$ excitations per bosonic mode $a_i$), this description provides a lower
bound on the performance of the cooling protocol, since in the absence of an
inhomogeneous Knight field cooling is limited to $\OO(1)$ excitations per mode
rather than the $\OO(\sqrt{N})$ coolable excitations expected at the beginning
of the cooling process for spins, cf.~\Eqref{HomCool}. In the two limiting
cases discussed so far, \Eqref{eqmgamma} simplifies to
\begin{equation*}
\frac{\Delta \gamma_{ij}}{\Delta t}=
\begin{cases}
-2  \kappa \delta_{ij} g_i^2\gamma_{ii}  & \text{($i$) Spin Temp.} \\
\xi_{ij} \gamma_{ij} -\kappa \sum_k g_k (g_i \gamma_{kj} + g_j
\gamma_{ik}) & \text{($ii$) Bosonic.}
\end{cases}
\end{equation*}

\begin{figure}
\begin{center}
\includegraphics[width=\columnwidth]{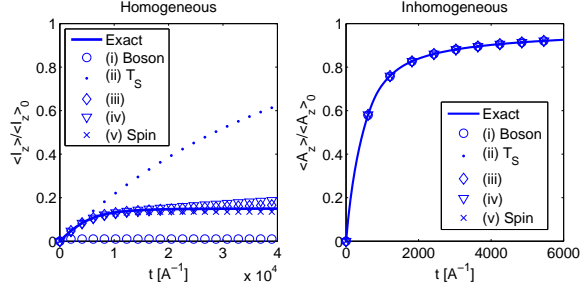}
 \caption{(Color online) Comparison of different approximation schemes for the
 homogeneous situation with $N=100$ (left) and the case of Gaussian
 couplings (as in Fig.~\ref{fig1})
 and $N=10$ nuclear spins (right). } \label{fig2}
\end{center}
\end{figure}

One can take into account more aspects of the spin algebra by
replacing some higher order expectation values by lower orders
using the properties of Pauli matrices $[\sigma_i^+,\sigma_i^-] = \sigma_i^z$
and $\sigma_i^z\sigma_i^\pm = \pm\sigma_i^\pm$, obtaining
\begin{align}
\frac{\Delta \gamma_{ij}}{\Delta t} & = \xi_{ij} \gamma_{ij} -
 \kappa \delta_{ij} \sum_k g_k (g_i \gamma_{kj} + g_j \gamma_{ik})
\nonumber\\ & -\kappa (1-\delta_{ij}) \Big( -  \sum_{k\neq i} g_k
g_i  \langle \sigma_k^+ \sigma_i^z \sigma_j^- \rangle  + g_i^2
\gamma_{ij} \nonumber\\ &\,\,\,\,\,\,\,\,\,\,\,\,\,\, -
\sum_{k\neq j} g_k g_j \langle \sigma_i^+ \sigma_j^z \sigma_k^-
\rangle + g_j^2 \gamma_{ij}\Big).
\end{align}
The remaining higher order expectation values (now having distinct
indices $i\neq j, j \neq k$) can be approximated in a Hartree-like
way~\cite{Aga71} ($iii$), or, having the bosonic limit in mind, by
the Wick theorem ($iv$),
\begin{equation*}
\frac{1}{2}\langle \sigma_k^+ \sigma_i^z \sigma_j^- \rangle=
\begin{cases}
(\gamma_{ii}-\frac{1}{2}) \gamma_{kj} &\text{($iii$)}, \\
-\frac{1}{2} \gamma_{kj} + \gamma_{ki}\gamma_{ij}+\gamma_{kj}\gamma_{ii}
&\text{($iv$).} \\
\end{cases}
\end{equation*}
The fifth and final approximation scheme we invoke has been
introduced in the context of superradiance as a Wick-type
factorization, that takes into account the partly bosonic, partly
fermionic properties of spin-1/2 operators~\cite{AEI93}. In
contrast to the last two factorization schemes, it does not rely
on distinction of cases. It is directly based on the exact
Eq.(\ref{eqmgamma}), and approximates the
three-operator-expectation values in the following way
\begin{equation*}
\frac{1}{2} \langle \sigma_k^+ \sigma_i^z \sigma_j^-
\rangle=-\frac{1}{2} \gamma_{kj} -
\gamma_{ki}\gamma_{ij}+\gamma_{kj}\gamma_{ii} \,\,\,\,\,\,\,\,\,
\text{($v$) ``Spin''.} \\
\end{equation*}

Direct comparison of the approximation schemes $(i)$--$(v)$ with
the exact solution for both homogeneous and inhomogeneous
couplings is shown in Fig.~\ref{fig2}. In the homogeneous case the
spin temperature description $(i)$ is clearly qualitatively wrong,
because it neglects correlations in the bath. The bosonic
description ($ii$) captures the feature of dark states, but it
overestimates their influence: Instead of $\sim \sqrt{N}$, only
one excitation can be removed. The two schemes based on
distinction of cases, $(iii)$ and $(iv)$, give very good results
initially, until roughly $\sqrt{N}$ spins have been flipped. Then
however, the polarization keeps increasing on a slow timescale and
does not reach a steady state in the correct time. The
$(v)$-``spin''-approximation gives very good results, and gets
both the polarization timescale and the finally obtained value of
the polarization right within a few percent.

The comparison of the different approaches to the exact solution
for inhomogeneous couplings is restricted to small particle
numbers (see Fig.~\ref{fig2}). In this regime all introduced
approximation schemes reproduce the exact dynamics correctly. The
reason for the good correspondence is the strong dephasing of dark
states and generally coherences between nuclear spins for small
inhomogeneous systems.

Using these approximations we present the polarization dynamics
for $N=10^3$ spins coupled through a 2D Gaussian wave function in
Fig.~\ref{fig3}. For the data presented in this and the following
figure, we considered the spins in a 2D square lattice geometry,
with the lattice constant set to unity. The bosonic description
displays the lowest final polarization and polarization rate (for
the same reasons as in the homogeneous case) and is expected to
give lower bounds on the performance on the polarization
procedure. Of particular interest are the predictions of the
($v$)-``spin''-approximation scheme, because its good performance
in the completely homogeneous situation gives confidence that also
partial homogeneities are correctly accounted for. Achieved
polarizations of $\sim 60 \%$ in this setting show the importance
of the intrinsic dephasing due to the inhomogeneity (homogeneous
coupling would allow for $<5\%$ polarization). However, the
intrinsic inhomogeneity alone does not allow for ultra-high
polarizations and we are thus lead to investigate more
sophisticated cooling schemes. As shown later, in these enhanced
protocols all approximation schemes lead to the same conclusions.

\begin{figure}
\begin{center}
\includegraphics[width=0.9\columnwidth,height=5cm]{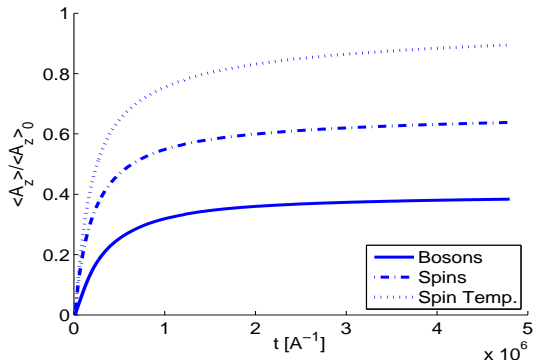}
  \caption{(Color online) The polarization dynamics for $N=1000$ spins coupled with
 a 2D Gaussian wave function, which is shifted from the origin by $1/3$ in
 $x$- and $y$-direction. } \label{fig3}
\end{center}
\end{figure}

To gain a better understanding of the presented phenomena in the
inhomogeneous situation, we go to an interaction picture $\rho_I =
U_0 \rho U_0^\dag$, with $U_0=\exp(-i {A}^z t/2)$, which shows
very clearly the oscillating coherences between spins with $g_i
\neq g_j$
\begin{align}
\label{iapic} \frac{\Delta \rho_I}{\Delta t} = -\kappa \Big[
\sum_{ij} g_i g_j e^{-ig(g_i-g_j)t/2} \sigma_i^+ \sigma_j^- ,
\rho_I \Big]_+ \nonumber\\ + 2\kappa \sum_{ij} g_i g_j
e^{-ig(g_i-g_j)t/2} \sigma_j^- \rho_I \sigma_i^+.
\end{align}
In the rotating wave approximation (RWA), the rotating terms
($g_i\neq g_j$) are neglected and in the absence of exact
symmetries the above equation reduces to the spin temperature
description. A partial rotating wave approximation neglects only
the coherences between spins with considerably different coupling
constants, i.e. the ratio between dephasing and polarization rate
is required to be large $(4 |g_i - g_j|/ (g \Delta t g_i g_j) >
1)$. This procedure gives a block diagonal Liouvillian which
allows for the extension of the numerical studies to particle
numbers up to $N=10^4$.

In the RWA we evaluate the build-up time $\tau_p$ for the
polarization as the inverse of the weighted average of the
individual spin decay times
\begin{equation}
\label{td1} \tau_p = \left(\frac{\sum_i g_i \kappa_i}{\sum_i g_i
}\right)^{\hspace{-3pt}-1}\hspace{-8pt}=\hspace{-1pt}\frac{4\sum_i
g_i}{g(g \Delta t)\sum_i g_i^3 }= \OO\left( \frac{4 N^{3/2}}{A (g
\Delta t)} \right),
\end{equation}
and find good agreement with the numerically obtained timescale to
reach the steady state in all discussed schemes. For example, for
the data presented in Fig.~\ref{fig3} we find times of
$3.4\times10^5$ (Spin Temp.), $4.6\times10^5$ (Bosonic), and
$3.3\times10^5$ (``Spin'') in units of $A^{-1}$ to reach
$(1-e^{-1})\approx 0.63$ of the quasi steady state
Overhauser-field. This agrees well with the analytical estimate
$\tau_p \approx 2.4 \times 10^5/A$; despite the differences in the
final polarizations obtained in the different approximation
schemes. This correspondence between the RWA-based estimate and
the numerically obtained polarization times for the coherent
evolution indicates that the inhomogeneous Knight field provides
coolable excitations at a rate larger than the polarization rate,
thus not slowing down the process.

When the inhomogeneity of the coupling is large enough to justify
the rotating wave approximation, each spin evolves with its own
Liouvillian and the nuclei remain in a product state during the
whole evolution. To keep the errors in the derivation of the master
equation (due to higher order terms of the expansion of the time
evolution operators in Eq.(\ref{BasicEvolution})) small, it is
sufficient to do so for each spin individually in this case. This
allows a larger time step $\Delta t \ll (A \alpha_{\rm max})^{-1}
=\OO(N/A) $ in each cycle and therefore the cooling rate can be
significantly enhanced. The cooling time effectively scales only
linearly in the particle number
\begin{equation}
\label{td2} \tilde{\tau}_p  =\OO\left( \frac{4N}{A (A /N \Delta
t)}\right) .
\end{equation}
Taking $A=100 \mu$eV$\sim 40$ps, a value typical for GaAs QDs, and
0.1 as the value for the terms $g \Delta t$ and $A/N \Delta t$ in
the denominators of Eqs.(\ref{td1}) and (\ref{td2}) respectively,
we find that approximately $4\times 10^3$ and $3 \times 10^5$
spins can be cooled to more than 90\% of the steady state value
$\langle A^z \rangle_{\rm ss}$ within a millisecond.

\begin{figure}
\begin{center}
\includegraphics[width=\columnwidth]{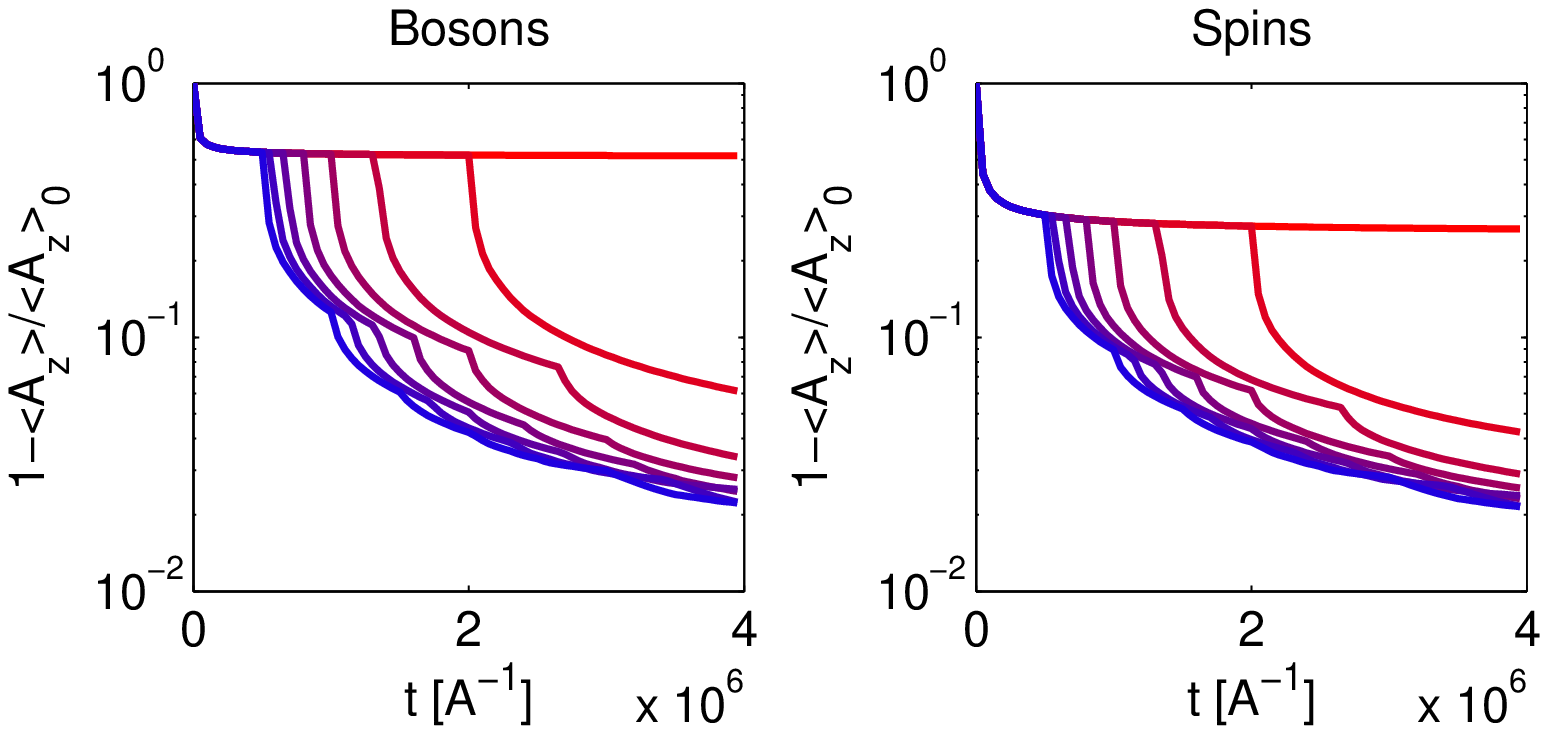}
\includegraphics[width=0.94\columnwidth,height=3.5cm]{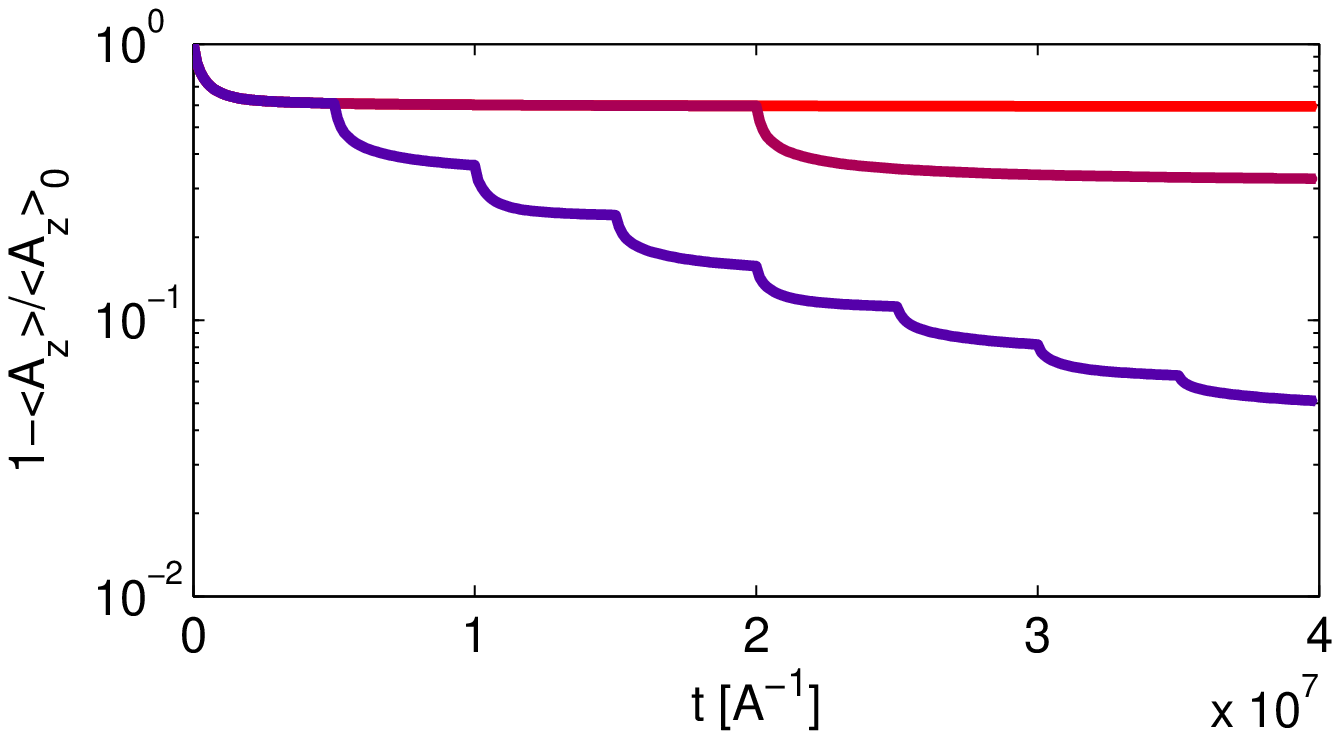}
\caption{(Color online) Polarization dynamics in the enhanced
cooling protocol
  for $N=196$ (upper plots) and $N=1000$ (lower plot). In the upper plots
  approximation schemes $(ii)$ (left) and $(v)$ (right) have been invoked, the
  lower plot is based on the bosonic model and the partial rotating wave approximation
  (see text). In all plots the different lines are representing
  cooling procedures
  with different numbers of modes changes. In the upper plots the randomly
  chosen Gaussian modes with width $w=N/4$ are defined by the centers
  $\{(1/3,1/3), (1.35,-0.81), (0.32,-0.04), (1.17,0.79),\newline
  (-0.13,-1.44), (0.96,-0.17), (0.35,0.88), (1.27,0.71)\}$. In the lower plot
  only two modes with centers $\{(1/3,1/3)$,$ (-3.15,-1.5)\}$ have been
  iterated.} \label{changemode}
\end{center}
\end{figure}

We now study enhanced cooling protocols that lift the dark-state
limitations and which rely solely on the ability to shift the
center of the electron wave function. These shifts can be effected
by applying dc gate voltages to the QD. After such a shift only
very few spins will have the same coupling constants for
\textit{both} wave functions and therefore singlet-like coherences
are broken up. We confirm this expectation numerically as shown in
Fig.~\ref{changemode} for some exemplarily chosen shifts of the
electron wave function. The shifts range from a few lattice sites
to roughly the width of the electron wave function. The timing of
the shifts we have performed for obtaining the data presented in
Fig.~\ref{changemode}, can be inferred from the plots, as it is
accompanied by a rapid increase in the cooling rate.

Regarding the approximation schemes, we have found that all schemes taking
into account coherences, ($ii$)-($v$), predict the same behavior, and the
spin-based factorization ($v$) offers the quantitatively best description.
It is important to note that all these descriptions coincide at the
end of the cooling protocol [shown in Fig.~\ref{changemode} only for ($ii$)
and ($v$)]. In particular the limiting bosonic model predicts the same high
($\ge 95 \%$) polarizations and cooling rates as the other schemes, which
leads us to conclude that $\OO(10)$ mode changes are sufficient to achieve
near-ground state cooling for realistically large numbers of nuclei in QDs.

Despite being a radical approximation at low polarization, the bosonic scheme
($ii$) captures the cooling dynamics qualitatively and we remark that it can
be generalized to provide an accurate and conceptually simple
description of the electron-nuclear spin dynamics at high polarizations
\cite{CGT+unpub}.

The cooling schemes we have presented are governed by the optimal timescale
set by the hyperfine interaction constant $A$, but the schemes themselves
leave room for optimization: The cooling rate can be tuned by choosing $\Delta
t$ adaptively during the cooling process. The mode changes can be optimized by
a careful choice of the size and the timing of the shifts, and through more
sophisticated deformations of the electron wave function. These and further
modifications are implementation-dependent and will be the topic of future
work.

In using the Hamiltonian \Eqref{Hamiltonian} we have neglected a
number of weak interactions that are present in actual systems
and, while being much smaller that the dominant hyperfine term,
may become important on the long time-scales required to reach
high polarization. We argue in the following that these terms do
not affect the quantitative conclusions obtained. While nuclear
Zeeman energies are large enough to cause additional dephasing
between the nuclear spins, similar to the inhomogeneous Knight
fields, this will only be effective between nuclei of
\emph{different} Zeeman energy, i.e., belonging to different
nuclear species. This leads to 2 to 3 mutually decohered subsystems
(in a partial rotating wave approximation) each of which is
described by our model.

The nuclear dipole-dipole interaction \cite{Sli80} can lead to
both diffusion and dephasing processes, both of which are of minor
importance as shown below. Dipolar processes that change $A^z$ are
off-resonant and hence expected to be slow, as indicated by the
nuclear spin diffusion rates measured, e.g., in \cite{Pag82} and
should not significantly affect the polarizations reached.
Resonant processes such as terms $\propto I_i^zI_j^z$ affect the
cooling process only insofar as they can cause dephasing of dark
states similar to the inhomogeneous Knight shift.  The rate at
which coolable excitations are provided is set by the energy
difference for two nuclear spins in a dark pair. The interaction
energy for two neighboring spins is about $\sim10^{-5}\mu$eV
\cite{SKL03}, hence a singlet of neighboring spins can dephase in
$\sim100\mu$s (or slower if all surrounding spins are polarized).
Even widely separated spins interacting with differently polarized
environments dephase only up to a few ten times faster than this
(depending on the geometry). Thus we see that the dipolar
dephasing is considerably slower than that caused by the
inhomogeneous Knight field and only if the latter becomes
inefficient due to homogeneities (towards the end of cooling a
given mode) the dipolar dephasing can contribute coolable
excitations, but at a much slower rate than what can be achieved
by changing the electron wave function and the ensuing return to a
situation of strong Knight inhomogeneity. Thus, one does not
expect the cooling process to be affected except for a slight
additional dephasing. However, on much longer timescales of $10$s
of ms the dipole-dipole interaction provides depolarizing
mechanism (affecting mainly nuclei with a weak hyperfine
interaction) that needs to be considered, e.g., when cooling much
beyond $90\%$ polarization is studied.

Clearly a polarization $<100\%$ of the electron ``reservoir''
directly translates into limitations on the final polarization of
the nuclei. A quantification of this necessarily needs to refer to
the details a concrete physical realization of our model, which is
not the topic of this article. The limitations can be minute, e.g.
in the case of the double dot setup presented in the next section.

\section{Adapting the model  to concrete physical settings}\label{sec:adapt}
The generic model of a single spin-1/2 particle coupled inhomogeneously to an
ensemble of $N$ nuclear spins can readily be adapted to various experimental
settings.

If a source of spin polarized electrons is available, single
electron tunneling into the QD provides the initialization.
Controlled tunneling into and out of the QD with rates
$>10\,\mr{ns}^{-1}$ appears feasible \cite{FHPD06,LA06},
justifying the description of the dynamics by a suddenly switched
on and off interaction.

For self-assembled QDs, optical pumping with polarized light has
been shown to provide a spin polarized bath of electrons that
cools the nuclei \cite{BSG+05,EKL+06,GOV+05,LMBI06,AFH06}.
However, in this setup the average dwell time of a single
polarized electron in the dot is large and the detuning due to the
$z$-component of the Overhauser field leads to instabilities
\cite{BUA+06,MLBI07,TWR+07} in the nuclear polarization which are
avoided in our scheme.

In double QDs in the two-electron regime~\cite{CoLo05,TPJ+06} the
role of the states $|\dn\rangle,|\up\rangle$ is played by the
two-electron singlet $|\tilde S\rangle$ and one of the triplet
states; in the following we consider
$|T_+\rangle=|\up\rangle|\up\rangle$. Tunnel coupling between the
two dots and the external magnetic field are chosen such that the
other triplet states are off-resonant and cause only small
corrections to the dynamics sketched here.

As discussed in more detail in \cite{ErNa02,CoLo05,TPJ+06} the
hyperfine interaction in this system is described by the
Hamilonian $\sum_l \bS_l\cdot\bA_l$, where $l=L,R$ refers to the
orbital state of the electron. Coupling between $|\tilde S\rangle$
and $|T_+\rangle$ is mediated by the \emph{difference} $\delta
A^\pm = (A_L^\pm-A_R^\pm)/2$ of the collective nuclear spin
operators of the two dots $L,R$, while the effective Overhauser
field is given by the sum $(A_L^z+A_R^z)/2$. Thus we have that the
analysis of the previous sections applies to the double dot case
in this regime (to zeroth order, cf. \cite{CoLo06}) with the
replacements
\begin{align} |\dn\rangle & \to |\tilde S\rangle, \,\,\,\,  |\up\rangle  \to
| T_+\rangle , \nonumber\\  A^\pm & \to - \sqrt{2} (\cos\theta)
\delta A^\pm, \,\,  A^z \to \frac{1}{2}(A_L^z+A_R^z).\nonumber
\end{align}
The adiabatic singlet has contributions from both the delocalized
$(1,1)$ and the localized $(0,2)$ charge states, and with $\cos
\theta$ we denote the amplitude of the $(1,1)$
contribution~\cite{TPJ+06} (with $(m,n)$ we denote a state with
$m$ electrons on the left and $n$ electrons on the right dot). The
effect of higher-order terms (e.g., of the nuclear spin components
$\delta A^z,A_L^\pm+A_R^\pm$) merits more detailed analysis.

This system is of particular interest since fast electrical
control of gate voltages can provide a highly spin polarized
electron system through near unit fidelity initialization of a
singlet in the right hand dot $|S(0,2)\rangle$
\cite{PJT+05,KBT+06}. Starting from this singlet, rapid adiabatic
passage ($1\,$ns~\cite{PJT+05}) by means of tuning the asymmetry
parameter $\epsilon$ between the dots, initializes the electrons
to the adiabatic singlet $|\tilde{S}\rangle$ and brings the system
to the $S-T_+$ resonance.

The transitions from the singlet to the other two triplets
$T_{0,-}$ are detuned by an external magnetic field (of order
$100\,$mT in the experiments of Ref.~\cite{PJT+05}). After a time
$\Delta t$ the system is ramped back to the (0,2) charge region
and the electrons relax to the singlet ground state, completing
one cooling cycle. If relaxation to the state $S(0,2)$ is fast,
the limiting timescale for this cycle is given by the hyperfine
coupling constant $A$, showing that here the polarization rate is
governed by the natural and optimal timescale (and not other,
slower timescales, like e.g.  cotunneling in
Refs.~\cite{OnTa04,RuLe06}).

In the GaAs double dot setup the sudden approximation is justified
for typical tunnel couplings $\sim 10 \mu {\rm eV}$, which have to
be compared to the typical timescale for a hyperfine flip $ \le
0.1 \mu$eV and the fact that additionally all spin flip
transitions are off-resonant during the adiabatic ramp. At the
$S-T_+$ resonance selecting a suitable combination of external
magnetic field and time step $\Delta t$ detunes the unwanted
transitions and at the same time ensures resonance for the
polarizing transition. Note also that the Overhauser field
increases the external magnetic field in materials with negative
electron $g$-factor, like GaAs ($g^\ast \approx -0.44)$, thus
further suppressing unwanted transitions and requiring retuning of
the end-point of the adiabatic ramp. Given the availability of
fast (100 ps) voltage pulses, the reinitialization of
$|S(0,2)\rangle$ via a $(0,1)$ charge state is likely to be
limited by the tunneling rate from the reservoir to the QD. For
optimal cooling efficiency this rate should and could be made
large $\gtrsim 10 A/ \sqrt{N}$~\cite{FHPD06,LA06}.

Since in the double dot setup the ``polarized'' state is a spin
singlet, there is no inhomogeneous Knight field to dephase the
dark states and DNP will be severely limited. However there are
many ways of providing it, for example by extending the cooling
cycle to include a third step in which a single-electron state of
the double dot is realized or by increasing the time spent at the
$S-T_+$ resonance in each cooling cycle (the latter would require
a reformulation of the master equation~(\ref{Meq}) not presented
here). At the same time it would be interesting to find evidence
for quantum coherence between nuclear spins in QDs by comparison
of the obtained Overhauser field in the case of strong and weak
inhomogeneous Knight fields~\footnote{Subradiance is not easily
demonstrated in quantum optical systems; it was experimentally
observed many years after superradiance, see
Ref.~\cite{PaCr+85}.}.

\section{Conclusions and Outlook}
In summary we have presented a quantum treatment of a dynamical nuclear spin
polarization scheme in single-electron quantum dots that takes into account
quantum coherences between nuclei and allows numerical study of the cooling
dynamics for thousands of spins. We have quantified limitations due to dark
states and shown that these limits are overcome by the inhomogeneous Knight
shift and active mode changes. From this we conclude that cooling to more than
$90\%$ (of the maximal Overhauser field) is feasible faster than typical
nuclear spin diffusion processes. Setups for the experimental realization of
our scheme have been proposed.

In order to go beyond the presented results to polarizations
larger than $99\%$, which would bring the system of coupled nuclei
close to a pure state and significantly reduce electron spin
decoherence, the presented scheme can be optimized, both in terms
of timing (length of the individual cooling step and wave function
changes) and in terms of the electron wave functions chosen. A
further enhancement may be achieved by combining the polarization
scheme with $A^z$-measurements \cite{SBGI06,KCL06,GTD+06} to
reduce the $A^z$ variance and to tailor the interaction times and
the external field to the measured $A^z$ value. Dipolar
interaction and other depolarizing processes will become more
important in later stages of the cooling and need to be considered
carefully in the development of ground-state cooling techniques.
More detailed studies of these processes may, in addition, lead to
schemes to monitor the intrinsic (dipolar) nuclear dynamics via
the hyperfine interaction.

The combination of high polarization and long coherence times make the nuclear
spin ensemble itself a candidate for an active role in quantum computation.
Like the actively explored single-nucleus-spin qubits~\cite{Kan98}, collective
excitations of a polarized ensemble of spins could also be used for quantum
information purposes~\cite{TGC+04}. Similar to their atomic
counterparts~\cite{KuPo03,DCZP00}, the ensembles might become more suited than
their isolated constituents for certain quantum information tasks.

\begin{acknowledgments}
We thank Bel\'en Paredes for very valuable discussions and Ata\c{c}
Imamo\u{g}lu for fruitful comments. This work was supported by
the DFG within SFB 631.
\end{acknowledgments}


\end{document}